\documentclass[aps,prl,twocolumn,groupedaddress,amssymb,amsmath,showpacs]{revtex4}
\usepackage{dcolumn}
\usepackage{graphicx}

\begin{document}


\title{Core-Level Photoelectron Spectroscopy Probing Local Strain at Silicon Surfaces and Interfaces}

\author{Oleg V. Yazyev}
\email[Electronic address: ]{oleg.yazyev@epfl.ch}
\affiliation{Ecole Polytechnique F\'ed\'erale de Lausanne (EPFL),
Institute of Chemical Sciences and Engineering, CH-1015 Lausanne, Switzerland}
\author{Alfredo Pasquarello}
\affiliation{Ecole Polytechnique F\'ed\'erale de Lausanne (EPFL),
Institute of Theoretical Physics, CH-1015 Lausanne, Switzerland}
\affiliation{Institut Romand de Recherche Num\'erique en Physique
des Mat\'eriaux (IRRMA), CH-1015 Lausanne, Switzerland}
\date{\today}

\begin{abstract}

Using a first-principles approach, we investigate the origin of
the fine structure in Si 2$p$ photoelectron spectra at the
Si(100)2$\times$1 surface and at the Si(100)-SiO$_2$ interface.
Calculated and measured shifts show very good agreement for both 
systems. By using maximally localized Wannier functions, we provide an 
interpretation in which the effects due to the electronegativity of 
second nearest neighbor atoms and due to the local strain field
are distinguished. Hence, in combination
with accurate modeling, photoelectron spectroscopy can provide a 
direct measure of the strain field at the atomic scale. 

\end{abstract}


\maketitle

The increasing availability of synchrotron radiation facilities
is bearing X-ray photoemission spectra of unprecedented resolution
characterizing surfaces and interfaces \cite{Himpsel90}. 
The achieved sensitivity is
sufficient to distinguish inequivalent subsurface atoms with
identically composed first-neighbor shells. Hence, the
interpretation of such core-level spectra can no longer be
achieved with simple electronegativity arguments, but requires
the consideration of the interplay between local strain fields 
and electronegativity effects of second-nearest neighbors. 

These difficulties are strikingly illustrated for the Si(100)2$\times$1 surface
and for the technologically relevant Si(100)-SiO$_2$ interface, 
which have been the objects of numerous highly resolved X-ray 
photoemission investigations. 
While the shifts pertaining to the first-layer dimer atoms 
of the Si(100)2$\times$1 surface have been identified, 
the other lines appearing in highly resolved Si $2p$ spectra still 
lack a consensual assignment \cite{Landemark92,DePadova98,Koh03}. For the Si(100)-SiO$_2$ interface,
highly resolved spectra show fine structure in
the nonoxidized Si line, with extra components at lower
(Si$^\alpha$) and higher binding energy (Si$^\beta$) with respect to the
Si bulk line \cite{Oh01,Dreiner05}.

We investigate the origin of the fine structure in Si $2p$ photoemission 
spectra at silicon surfaces and interfaces using density functional theory 
calculations based on pseudopotentials (PPs) and plane waves \cite{Car85-CPMD}.
To interpret the Si 2$p$ photoelectron spectra at silicon 
surfaces and interfaces, it is necessary to evaluate core-level 
shifts with respect to the Si bulk line. 
Since these shifts mainly result from the
relaxation of valence electron states, their accurate
determination is possible within a PP scheme, which does not
treat core electrons explicitly \cite{Pehlke93,Pasquarello95}. 
We calculated Si 2$p$
shifts including the effect of core-hole relaxation by taking 
total energy differences between two separate self-consistent calculations.
First, the ground-state energy is determined; then the PP of a
given Si atom is replaced by another PP which simulates the
presence of a screened $2p$ hole in its core \cite{Pehlke93}.

\begin{table}
\caption{\label{tab2} Comparison between calculated and measured
Si $2p$ shifts at the Si(001)-$c$(4$\times$2) surface. The shifts
(in eV) are given with respect the Si bulk line. The meaning of labels 
is explained in Fig.~\ref{fig1}.}
\begin{ruledtabular}
\begin{tabular}{ccccc}
   & \multicolumn{1}{c}{Theory}  &  \multicolumn{3}{c}{Experiment} \\
   & \multicolumn{1}{c}{Present} &
\multicolumn{1}{c}{Ref.\ \cite{Landemark92}} &
\multicolumn{1}{c}{Ref.\ \cite{DePadova98}}  &
\multicolumn{1}{c}{Ref.\ \cite{Koh03}}\\
\hline
 $1u$  &  -0.49 &  -0.49   &  -0.49  &   -0.50  \\
 $1d$  &   0.02 &   0.06   &   0.06  &    0.06  \\
 $2$   &  -0.01 &          &         &          \\
 $3$   &  -0.14 &  -0.21   &  -0.20  &   -0.21  \\
 $4$   &  -0.26 &          &         &          \\
 $3'$  &   0.24 &   0.22   &   0.20  &    0.23  \\
 $4'u$ &   0.21 &          &         &          \\
 $4'd$ &   0.10 &          &         &          \\ 
\end{tabular}
\end{ruledtabular}
\end{table}

\begin{figure}
\includegraphics[width=3.0cm]{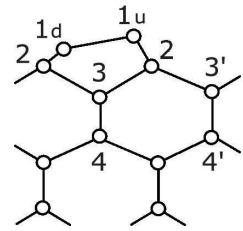}
\caption{\label{fig1}
The adopted labeling of Si atoms at the Si(001)-$c$(4$\times$2) surface. 
The $4'$ atoms can be further 
separated into inequivalent up ($4'u$) and down ($4'd$) atoms.}
\end{figure}

For the Si(100)-c(4$\times$2) 
surface we find binding energies of $-$0.49 eV and 0.02 eV for 
dimer-up and dimer-down atoms, in good agreement with experimental results \cite{Landemark92,DePadova98,Koh03}. 
The second-layer
atoms are found to yield very small shifts ($-$0.01~eV) while 
third- and fourth-layer atoms give shifts to
both lower ($3$, $4$) and higher binding energies ($3'$, $4'$) with
respect to the Si bulk line (Table~\ref{tab2}). Shifts of deeper layers are
negligible.

\begin{figure}
\includegraphics[width=8.5cm]{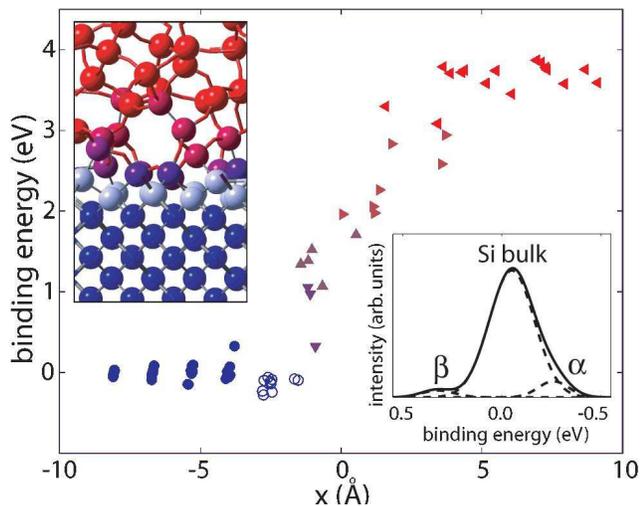}
\caption{\label{fig2}
Calculated Si 2$p$ shifts at the Si(100)-SiO$_2$
interface along an orthogonal direction to the interface.
The Si$^{+n}$ correspond to oxidized Si atoms with $n$
O neighbors ($n$=1 - $\blacktriangledown$, 2 - $\blacktriangle$, 
3 - $\blacktriangleright$, 4 - $\blacktriangleleft$) and are 
color-coded in the atomistic model. 
Nonoxidized Si atoms (Si$^0$) with ($\circ$) and without ($\bullet$) 
second-neighbor O atoms are distinguished. 
Inset: simulated spectrum for Si$^0$ 
(Gaussian broadened: $\sigma$=0.08~eV) and
its decomposition in three components.}
\end{figure}

For the Si(100)-SiO$_2$ interface, we adopt a model structure which 
incorporates a realistic transition region \cite{Bongiorno03}.
Calculated binding energies are
given in Fig.\ \ref{fig2} along the direction orthogonal to the
interface plane. The binding energies of oxidized Si atoms
increase almost linearly with oxidation state
\cite{Pasquarello95}, and show quantitative agreement
with experimental values (Table \ref{tab3}). The focus of the
present investigation is on nonoxidized Si atoms, which give
shifts with a significant spread near the interface. The
simulated spectrum associated to these Si atoms could be
decomposed into three components (Fig.\ \ref{fig2}, inset),
yielding shifts for the Si$^\alpha$ and Si$^\beta$ lines of $-$0.21 eV and 0.32 eV,
which also agree with experimental data (Table~\ref{tab3}).  

\begin{table}
\caption{\label{tab3} Comparison between calculated and measured
Si $2p$ shifts at the Si(100)-SiO$_2$ interface. The Si bulk line
is taken as reference.}
\begin{ruledtabular}
\begin{tabular}{ccccc}
   & \multicolumn{1}{c}{Theory}  & \multicolumn{2}{c}{Experiment} \\
   & \multicolumn{1}{c}{Present} &
                   \multicolumn{1}{c}{Ref.\ \cite{Oh01}}     &
                   \multicolumn{1}{c}{Ref.\ \cite{Dreiner05}}\\
\hline
 $\alpha$  &   -0.21 &  -0.25 &    -0.22 \\
 $\beta$   &    0.32 &   0.20 &     0.34 \\
 +1        &    0.78 &   1.00 &     0.95 \\
 +2        &    1.40 &   1.82 &     1.78 \\
 +3        &    2.37 &   2.62 &     2.60 \\
 +4        &    3.64 &   3.67 &     3.72 \\
\end{tabular}
\end{ruledtabular}
\end{table}

Binding energy shifts of Si atoms with identically composed first-neighbor shells 
are caused by electron density displacements due to either the local strain fields 
or electronegativity effects.
We carried out an analysis in terms of maximally localized Wannier functions (MLWFs) 
\cite{Marzari97}
in order to distinguish between these two effects \cite{Yazyev06}.
For the Si(100)-c(4$\times$2) surface, we find that, apart from the shifts associated
with the dimer atoms, the other shifts mainly result from the local strain induced 
by the surface reconstruction.
For the interface, the Si atoms with second-neighbor O atoms contribute to the
Si$^\alpha$ line, while the Si$^\beta$ line originates from bond elongations 
of Si atoms without second-neighbor O atoms.  From the experimental shift of about 
0.3~eV, we infer the occurrence of Si atoms with an average bond length elongation 
of $\sim$~0.05~\AA.

In conclusion, we revealed the physical mechanisms underlying 
the fine structure in Si 2$p$ photoelectron spectra 
at silicon surfaces and interfaces. 
Maximally localized Wannier functions offer a powerful tool 
to identify core-level shifts originating from the
electronegativity of farther neighbors. A key result of our work
is then that the remaining lines provide an atomic-scale probe 
of the strain in the structure. This confers to photoelectron 
spectroscopy a new functionality in addition to the detection
of chemical composition.


\begin{thebibliography}{12}

\bibitem{Himpsel90}
F. J. Himpsel {\it et al.},
in {\it Proceedings of the International School of Physics Enrico Fermi},
edited by M. Campagna and R. Rosei (Elsevier, Amsterdam, 1990), p.\ 203--236.

\bibitem{Landemark92}
E. Landemark {\it et al.},
{Phys.\ Rev.\ Lett.}\ {\bf 69}, 1588 (1992).
  
\bibitem{DePadova98}
P. De~Padova {\it et al.},
{Phys.\ Rev.\ Lett.}\ {\bf 81}, 2320 (1998).

\bibitem{Koh03}
H. Koh {\it et al.},
{Phys.\ Rev.\ B} {\bf 67}, 073306 (2003).

\bibitem{Oh01}
J.~H. Oh {\it et al.},
{Phys.\ Rev.\ B} {\bf 63}, 205310 (2001).

\bibitem{Dreiner05}
S. Dreiner, M. Sch\"urmann, and C. Westphal,
{J.\ Electron Spectrosc.\ Relat.\ Phenom.}\ {\bf 144-147}, 405 (2005).

\bibitem{Car85-CPMD}
R. Car and M. Parrinello,
{Phys.\ Rev.\ Lett.}\ {\bf 55}, 2471 (1985);
CPMD version 3.9.1, Copyright IBM Corp 1990-2004,
Copyright MPI f\"ur Festk\"orperforschung
Stuttgart 1997-2001, \texttt{http://www.cpmd.org}.

\bibitem{Pehlke93}
E. Pehlke and M. Scheffler,
{ Phys.\ Rev.\ Lett.}\ {\bf 71}, 2338 (1993).

\bibitem{Pasquarello95}
A. Pasquarello, M.~S. Hybertsen, and R. Car,
{Phys.\ Rev.\ Lett.}\  {\bf 64}, 1024 (1995);
{ Phys.\ Rev.\ B} {\bf 53}, 10942 (1996).

\bibitem{Bongiorno03}
A. Bongiorno, A. Pasquarello, M.~S. Hybertsen, and L. C. Feldman,
{ Phys.\ Rev.\ Lett.}\ {\bf 90}, 186101 (2003).

\bibitem{Marzari97}
N. Marzari and D. Vanderbilt,
{ Phys.\ Rev.\ B} {\bf 56}, 12847 (1997).

\bibitem{Yazyev06}
O.~V.~Yazyev and A.~Pasquarello,
{ Phys.\ Rev.\ Lett.} {\bf 96}, 157601 (2006).

\end{thebibliography}
\end{document}